# Antiferromagnetic cavity magnon polaritons in collinear and canted phases of hematite

I. Boventer*[1], H. T. Simensen[2], B. Brekke[2], M. Weides[3], A. Anane[1], M. Kläui[2,4,5], A. Brataas[2] and R. Lebrun*[1]

1. *Unité Mixte de Physique CNRS, Thales, Université Paris-Saclay, Palaiseau 91767, France*
2. *Center for Quantum Spintronics, Department of Physics, Norwegian University of Science and Technology, Trondheim, Norway*
3. *James Watt School of Engineering, Electronics & Nanoscale Engineering Division, University of Glasgow, Glasgow G12 8QQ, United Kingdom*
4. *Institut für Physik, Johannes-Gutenberg-Universität Mainz, D-55099, Mainz, Germany*
5. *Graduate School of Excellence Materials Science in Mainz (MAINZ), Staudinger Weg 9, D-55128, Mainz, Germany*

**Cavity spintronics explores light matter interactions at the interface between spintronic and quantum phenomena. Until now, studies have focused on the hybridization between ferromagnets and cavity photons. In this article, we realize antiferromagnetic cavity-magnon polaritons. The collective spin motion in single hematite crystals (α-$Fe_2O_3$) hybridizes with 18 – 45 GHz microwave cavity photons with required specific symmetries. We show theoretically and experimentally that the photon-magnon coupling in the collinear phase is mediated by the dynamical Néel vector and the weak magnetic moment in the canted phase by measuring across the Morin transition. The coupling strength $\tilde{g}$ is shown to scale with the anisotropy field in the collinear phase and with the Dzyaloshinskii-Moriya field in the canted phase. We achieve a strong coupling regime both in canted (C > 25 at 300 K) and noncolinear phases (C > 4 at 150 K) and thus coherent information exchange with antiferromagnets These results evidence a generic strategy to achieve cavity-magnon polaritons in antiferromagnets for different symmetries, opening the field of cavity spintronics to antiferromagnetic materials.**

The generation [1]–[4] and the manipulation [5], [6] of cavity-magnon polaritons (CMP) and the exploring strategies to integrate them into devices is at the core of cavity spintronics. The CMP is the quasiparticle associated with the hybridization of a cavity photon and a magnon, the quanta of a spin-wave [7]. Research directions in cavity magnonics range from studies of quantum information processing [8]–[11] to long-distance transport of spin current via cavities [12] and optomagnonics [13]–[15]. The hybrid nature of a CMP allows for the interconnection with (nonlinear) physical systems such as superconducting qubits, leading to quantum magnonics and the possibility of quantum-based information processing resources such as quantum memories [16]–[18]. Cavity photons couple coherently or dissipatively to magnons [5], [6], [19], [20]. A coherent transfer of information [16], [19]–[22] is achieved when $\tilde{g} \gg \kappa_c, \kappa_m$, that is, when the macroscopic coupling strength $\tilde{g}$ exceeds the dissipation parameters (the cavity resonator linewidth $\kappa_c$ and the magnon linewidth $\kappa_m$) with the experimental hallmark of observing an avoided level crossing.

Cavity magnonics has mainly focused on ferromagnetic materials, and especially on Yttrium-Iron-Garnet (YIG), which exhibits the lowest known Gilbert damping in ferromagnetic materials [16], [23]–[25]. Antiferromagnetic materials possess higher frequency modes that can reach the THz range and have in recent years been at the heart of intense research in spintronics [26]–[28]. Furthermore, AFM materials can exhibit various magnetic configurations, ranging from collinear easy-axis and easy-plane ordering to non-collinear spin-textures. Recent research highlights that insulating antiferromagnets can also possess ultra-low magnetic damping leading to long distance spin transport [29]–[33]. However, antiferromagnetic materials often couple weakly to external excitations due to vanishing stray fields. Thus, the hybridization of AFM magnons with cavity photons remains largely unexplored, with only a few recent solely experimental reports at millikelvin and sub-THz frequencies [34]–[38] which do not study the nature of the coupling strength in detail. The development of antiferromagnetic cavity magnonics thus requires solving two key bottlenecks. On the fundamental side, it requires an understanding of how cavity photons can efficiently couple with the dynamics of the Néel order in compensated AFMs, and in canted

---

* Corresponding authors: isabella.boventer@cnrs-thales.fr, romain.lebrun@cnrs-thales.fr

AFMs, potentially with their weak canted moments. On the applied side, it requires a demonstration of an easily accessible strong coupling regime. In turn, the latter would enable the incorporation of other nonlinear components for quantum cavity magnonics and future downscaling to on-chip platforms.

This article explores antiferromagnetic cavity photon magnon polaritons (A-CMPs) and their associated dynamics in collinear and non-collinear antiferromagnets. We use the antiferromagnet hematite ($\alpha - Fe_2O_3$) below and above the Morin transition. We demonstrate both theoretically and experimentally that the coupling strength scales as $\tilde{g}(T) \propto (H_{A'}(T)/8H_E)^{1/4}$ [39] in the collinear phase and $\tilde{g} \propto \sqrt{(H_D + H_0)^2/\omega_0 H_E}$ in the non-collinear phase ( with $H_{A'}$: the easy-axis anisotropy field, $H_E$: the exchange field, $H_D$: the Dzyaloshinskii-Moriya field and $H_0$: the applied magnetic field). We demonstrate the creation of A-CMPs for both the collective spin motion associated with the right-handed low-frequency mode in the canted easy-plane (cEP) phase (at T = 300 K), and with the left-handed mode below the Morin transition in the collinear easy axis (EA) phase, coupled with the transverse microwave field of a high-frequency cavity resonator operating in the 18 – 45 GHz range. Due to the low magnetic damping of hematite $\kappa_m$ and a low cavity linewidth $\kappa_c$, we reach a cooperativity $C = \tilde{g}^2/(\kappa_c \kappa_m) > 1$, both in the collinear (C > 4) and canted phase (C > 25) and for different cavity modes. Our findings demonstrate a persistent coherent information exchange between cavity resonator photons and antiferromagnetic magnons for all temperatures and two different AFM configurations.

We first model the coupling of a cavity resonator mode with an AFM magnon using the Hamiltonian of the entire system $\mathcal{H}_{sys} = \mathcal{H}_c + \mathcal{H}_{afm} + \mathcal{H}_{c-afm}$ with $\mathcal{H}_c$, $\mathcal{H}_{afm}$, $\mathcal{H}_{c-afm}$ describing the cavity photons, the antiferromagnetic magnons and the cavity photon-magnon coupling, respectively. Specifically (cf. SM.1), :

$$\mathcal{H}_{sys} = \hbar\omega_c \, aa^\dagger + \hbar\omega_m mm^\dagger + \hbar \, \tilde{g} \left(m^\dagger a + a^\dagger m\right), \quad (2)$$

where $\omega_c$, $\omega_m$ correspond to the respective frequencies of the cavity photon and AFM magnon modes and $\tilde{g}$ quantifies the macroscopic photon-magnon interaction, i.e., denotes the macroscopic coupling strength of the cavity photons with the AFM magnons. Importantly, $\tilde{g}$ scales as $g_0\sqrt{2NS}$ with $g_0 = \frac{\eta|\gamma|}{2}\sqrt{\frac{\hbar\omega_0 \mu_0}{2V_m}}$ being the single spin coupling strength ($\omega_0$ is the resonance frequency of the polariton where $\omega_c = \omega_m \equiv \omega_0$, $V_m$ denotes the cavity mode volume. The dimensionless factor $\eta$ accounts for the (spatial) mode overlap between the cavity photons and the magnons and *NS* corresponds to the total spin numbers involved in the coupling [41]. In AFMs, the expression of $\tilde{g}$ also depends on the AFM configurations (easy-axis and easy-plane) and requires distinguishing between collinear and non-collinear cases. In **Table 1**, we present the derived effective coupling strength for both collinear easy-axis AFMs [39] and canted antiferromagnets (see **SM.1** [40]):

| Magnetic ordering | Collinear (easy axis) AFM ($H < H_{sf}$) | Non-collinear (canted) AFM |
|---|---|---|
| Frequencies [33] [41] | $\omega_{0,\pm} = \gamma\sqrt{2H_E H_{A'}} \pm \gamma H_0$ (3a) | $\omega_{0,+} = \gamma\sqrt{2H_E(H_{A'} + H_a) + H_D(H_D + H_0)}$ (3b) <br> $\omega_{0,-} = \gamma\sqrt{2H_E H_{A'} + H_0(H_0 + H_D)}$ (3c) |
| Coupling strength [39] | $\tilde{g}_{EA} = g' \cdot \left(\frac{H_{A'}}{8H_E}\right)^{1/4}$ (4a) <br> with $g' = g_0\sqrt{2NS}$ | $\begin{cases} \tilde{g}_{x,z,cEP} = -ig'\sqrt{\dfrac{2\gamma(H_0 + H_D)^2}{\omega_0 H_E}} \text{ (4b)} \\ \tilde{g}_{x,y,cEP} = g'\sqrt{\dfrac{2\omega_0}{\gamma H_E}} \text{ (4c)} \end{cases}$ |

*Table 1: Summary of the AFM resonance frequencies $\omega_{0,\pm}$ and of the coupling strength $\tilde{g}$ associated with A-CMPs for collinear easy-axis ($H < H_{sf}$: the spin flop field) and non-collinear canted AFMs where $H_E$ denotes the exchange field. For collinear easy-axis AFMs, $H_{A'}$ stands for the easy-axis anisotropy field and $H_0$ for the magnetic field applied along the easy-axis. For the noncollinear AFMs, $H_0$ stands for the magnetic field applied along the canted moment. For the canted easy-plane case, $H_a$ stands for the hard-axis anisotropy and $H_{A'}$ the weak easy-axis anisotropy within the easy-plane. In contrast, setting $H_a = 0$, yields a description of canted easy axis antiferromagnets. The expressions of the*

*coupling strength show that it is independent with the applied magnetic field $H_0$ in the collinear easy-axis case, and that the coupling strength depends on the magnetic field and on the polarization of the cavity photons in the non-collinear canted case. The subscript $x,y$ $(z)$ denotes the photon mode that propagates in the $x$ direction with an electric field polarized in the $y$ $(z)$ direction. Here the $y$ axis is the easy - axis.*

For collinear easy-axis AFMs, A-CMPs originate from the strong coupling of the cavity photons with the dynamics $\left(\propto \left(\frac{H_{A'}}{8H_E}\right)^{\frac{1}{4}}\right)$ of the compensated Néel vector (**Eq. 4a**). For collinear easy-plane AFM, the coupling is zero in absence of externally applied magnetic fields. For non-collinear AFMs, A-CMPs emerge from the coupling with the dynamics of both the canted net magnetic moment $\left(\propto \sqrt{\frac{(H_D+H_0)^2}{\omega_0 H_E}}\right)$ or the Néel vector $\left(\propto \sqrt{\frac{2\omega_0}{\gamma H_E}}\right)$ depending on the photon polarization (**Eq. 4b-c**). Depending on the strength of the DMI field and on the magnon gap, different strategies can thus be developed to reach the strong coupling regime.

We then turn towards the experimental observations of antiferromagnetic cavity magnon-polaritons using single crystals of hematite ($\alpha$-Fe$_2$O$_3$, volume: 1.25mm$^3$). We place the sample in a 3D millimeter-sized, rectangular microwave cavity (**Figure 1 (a)**) with cavity resonances from 18 – 45 GHz (see **SM.2** [40]). As schematically shown in **Figure 1 (a)**, we record the reflective absorption spectrum of our hybrid system with a vector network analyzer (VNA) while sweeping the external magnetic field $H_0$ over the anticrossing(s). At $T = T_{Morin}$ ($\sim 260\,K$ [42]), the magnetic configuration of hematite changes from a canted easy-plane (cEP) (induced by a bulk Dzyaloshinskii-Moriya interaction (DMI)) antiferromagnet to an easy-axis (EA) collinear AFM for $T < T_{Morin}$. Hematite also possesses AFM modes accessible at tens of GHz (**Figure 1 (c)**): for $T > T_{Morin}$, a right-handed low-frequency mode (with a gap $\approx$ 15 GHz, due to the small in-plane anisotropy) ; for $T < T_{Morin}$, a high-frequency left-handed AFM mode that can be softened to GHz frequencies with an externally applied field (due to a spin-flop field of less than 8 T [43]). The corresponding orientation of the magnetic sublattices for these two modes is shown in **Figure 1 (c & e)**. This variety of AFM modes renders hematite a well-suited model system for investigating A-CMPs in collinear and canted AFMs and to generalize the existence of A-CMPs in these two different types of AFMs. To maximize the coupling between the cavity field and the magnons, we choose the polarization of the photon propagation to be perpendicular to the easy axis in the EA phase (**Eq. 4a**) and the canted moment in the cEP phase (**Eq. 4b**). Note, that $\sqrt{\frac{2\gamma(H_D+H_0)^2}{\omega_0 H_E}} > \left(\frac{H_{A'}}{8H_E}\right)^{\frac{1}{4}}$ which implies that the coupling is larger in the cEP phase as compared to the EA phase. The presence of avoided level crossing (anticrossing) with a frequency gap $\Delta\omega_{gap} = 2\tilde{g}$ at resonant coupling conditions ($\omega_c = \omega_m = \omega_0$) represents the hallmark of strongly coupled cavity photon-magnon states, that is an A - CMP. We observe clear evidence of such anticrossings in both phases, as shown in **Figure 1 (b)** for T = 50 K and **(d)** for 300 K. As expected, there is a decrease in the resonance frequency with the applied field for the left-handed mode in the collinear EA phase and an increase for the right-handed mode in the non-collinear cEP phase.

Determining the specific coupling regime and the magnon-photon cooperativity C requires an analysis of the coupling strength and cavity photon- and magnon linewidths. To extract the coupling strength, we first fit the anticrossings following the field-dependent dispersion for the lowest modes (cf. **Table 1**) in the EA and cEP phases ([40], **SM.3 (b)**). Then, we measure across the Morin transition and display in **Figure 2 (a)-(b)** the temperature dependence of the coupling strength for a cavity resonance of $\frac{\omega_{0,-}}{2\pi} =$ 20.8 GHz for the EA phase and 27.5 GHz for the cEP phase. Note, that the exciting AC fields of the cavity need to be orthogonal to the Néel vector or canted moment in the two respective antiferromagnetic phases. Due to the rotation of the sublattices by 90° at the Moring transition, this condition is fulfilled for different cavity modes in the two phases. The coupling modes with lowest resonance frequencies are at around 20 GHz for T < T$_{Morin}$ and 27 GHz for T > T$_{Morin}$. Beyond the symmetry of the AC fields, the identification of the true cavity modes relies on identifying the mode with phase jump response of $\pi$ (cf. supplementary material S2 (c)) for more details). The coupling strength reaches hundreds of MHz in the two phases, with larger values in the canted phase, demonstrating that one can stabilize A-CMPs through the coupling between cavity photons and the weak canted moment of AFMs. The larger coupling strength in the EP phase agrees with the theory (**Eq. (4)**) as $\sqrt{\frac{2\gamma(H_D+H_0)^2}{\omega_0 H_E}} / \left(\frac{H_{A'}}{8H_E}\right)^{\frac{1}{4}} \approx 10$ (for T ≈ 100 K). The value $H_D \approx 2.2$ T is extracted from SQUID measurements and $H_{A'}$ from the resonance frequencies.

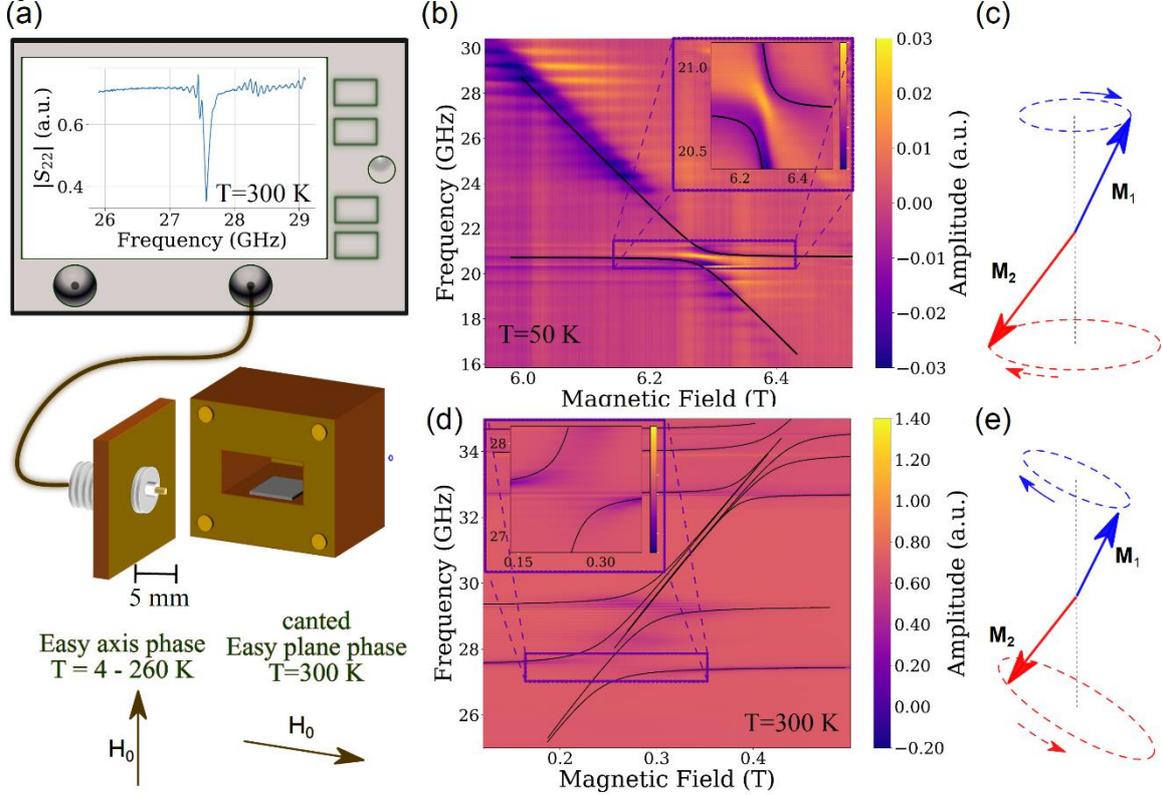

*Figure 1: **Cavity photon-magnon polaritons in the collinear and canted phase of hematite** (a) Schematic of the experimental setup employing reflective vector network analysis. Using hematite, we performed temperature-dependent measurements for antiferromagnets with different symmetries, in the easy-axis (T = 50 - 260 K) and the canted easy-plane phase T = 300 K). The external magnetic field $H_0$ is aligned with the easy axis in the collinear phase (a-axis sample) and perpendicular to the hard axis (c-axis sample) in the easy-plane phase of hematite. The 'VNA-image' shows the data for zero field at 300 K. (b) Exemplary spectrum of the reflection measurement ($S_{22}$) including fits according to the theory (black solid lines) at T = 50 K. The inset shows a zoom of an avoided crossing and a fit with $\omega_0/2\pi = 20.8$ GHz and $\tilde{g}_{EA}/2\pi = 400$ MHz (c) Sublattice orientation of the left-handed mode of hematite in the easy axis phase. (d) Reflective amplitude spectrum of the A-CMP in the cEP phase at T = 300 K including fits according to the theory (black solid lines). Inset: Zoom including the fit to a selected avoided crossing at $\omega_0/2\pi = 27.5$ GHz and $\tilde{g}_{cEP}/2\pi \approx 680$ MHz. (e) Sublattice orientation of the right-handed mode of hematite in the easy-plane phase. The sample size is 5mm x 5mm x 0.5 mm for both phases.*

In the collinear phase, the coupling strength is largest at low temperature and decreases with temperature in agreement with the decrease in magnetic anisotropy [43]. Using COMSOL® Multiphysics (cf. [40], **SM.4**) to simulate the cavity modes, we establish a more quantitative comparison between theory and experiment by determining $g' = g_0\sqrt{2NS}$. We obtain $g_0/2\pi \approx 120$ mHz and thus theoretical values of $\tilde{g}$ (green dotted line in **Figure 2 (a)**), which agree with the experiment. Further, **Figure 2 (a)** shows the temperature dependence of the effective macroscopic coupling strength for a resonance frequency of $\frac{\omega_{0,-}}{2\pi} = 20.8$ GHz for the EA phase and 27.5 GHz for the cEP phase, including the temperature dependence of the resonance field and the model based on Eq. 4 and Ref. [43]. In **Figure 2 (d),** we display the corresponding spectra for 50 K – 180 K. One can see the collapse of the coupling strength for temperatures above 150 K, corroborating the results from **Figure 2 (a).** In conclusion, the coupling strength highlights the role of the DMI and anisotropy fields in the canted and collinear phases, respectively, as expected from the theory. Regarding the additional increase of the coupling strength at lower temperatures, it could arise from changes of the dielectric constant of hematite which is out of the scope of this paper focusing on the investigation of the generic existence of A-CMPs.

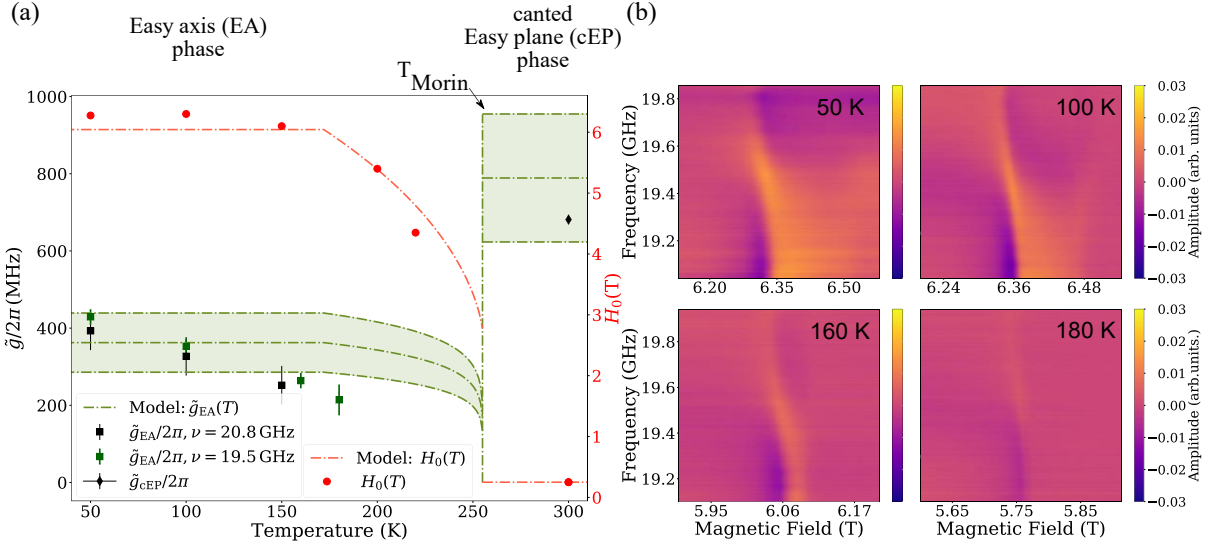

*Figure 2: **Effective coupling strength of antiferromagnetic cavity photon-magnon polaritons (A-CMP).** (a) Resonant fields (red circles) and effective coupling strengths (black and green squares) as a function of temperature. Theoretical expectations are shown in dashed lines. Two independent datasets are measured for frequencies of 20.8 GHz (black squares) and 19.5 GHz (green squares). Towards the Morin-Transition (T ~ 260 K), the coupling strength decreases due to the decreasing easy-axis anisotropy field $H_{A'}$. Correspondingly, for $T > T_{Morin}$, the coupling strength is in agreement with the theoretical value within error bars (green shaded) (**Eq. 4**). (b) Anticrossing spectra at different temperatures for a resonance frequency of 19.5 GHz, showing the collapse of the coupling strength in the collinear phase, towards the phase transition at $T_{Morin}$. The VNA output power level was set to - 6 dBm. Note, that the frequencies at 20.8 GHz and 19.5 GHz correspond to the same mode but the resonance frequencies slightly shifted due small displacement of the sample position between measurements (cf. supplementary material S5)*

We finally quantify the degree of coherent information exchange between AFM magnons and photons via the cooperativity $C = \tilde{g}^2/(\kappa_c \kappa_m)$. We extract the cavity resonator's linewidth (see **Figure 3 (a)**, green squares) for each mode and the magnon linewidth (blue squares) from off-resonant VNA measurements (see **SM. 3 (d-e)** [40]). We also confirm the magnon linewidth from conventional stripline ferromagnetic resonance measurements (see **SM. 3** [40]). As shown in **Figure 3 (a)**, we measure a total decrease of the magnon linewidth from 50 K to room temperature (300 K) across the Morin transition in agreement with previous reports (cf. Ref. [43]). From the values of $\tilde{g}$, $\kappa_c$ and $\kappa_m$, we then determine the cooperativity of the hybridized states as displayed in **Figure 3 (b).**

For T < 150 K, we achieve C > 1 (green dotted line), exceeding C > 4 at T = 150 K. Thus, even in the absence of a net static magnetic moment, the coupling between the Néel order dynamics and the cavity field is strong enough to enable a coherent information transfer. This is the first crucial step to couple the A-CMP to other (nonlinear) systems [16], [39]. In the canted phase, we achieve C > 25, which is far above C = 1 highlighting the potential of canted AFM for cavity spintronics. Finally, we determine the coupling strength for different sample volumes (2.5 to 12.5 mm³). The sample volume is of the order of the cavity mode volume. Hence, the sample modifies the cavity resonance frequencies ([40], **SM.2 (b)**). From the anticrossings observed at different resonance frequencies $\omega_0$, we can, however, determine the coupling strength for different sample sizes (inset **Figure 3 (b)**).

As expected, the coupling strength scales with the number of contributing spins originating from the weak canted moment, i.e., the relative spin number $N/N_0$, where $N_0$ is the spin number of the largest sample (black squares, dotted). Considering the frequency shift induced by the sample volume ([40], **SM. 2**), the coupling strength follows $\tilde{g} \propto \sqrt{\frac{N}{\omega_0}}$ (cf. Eq. (4)). The prevailing existence of A-CMPs for different sample sizes represents another key feature to open new research directions in cavity magnonics. Specifically, orthoferrites [44]–[49] with large DMI fields and sub-THz resonance frequencies could generate A-CMPs for a different symmetry of photons polarizations.

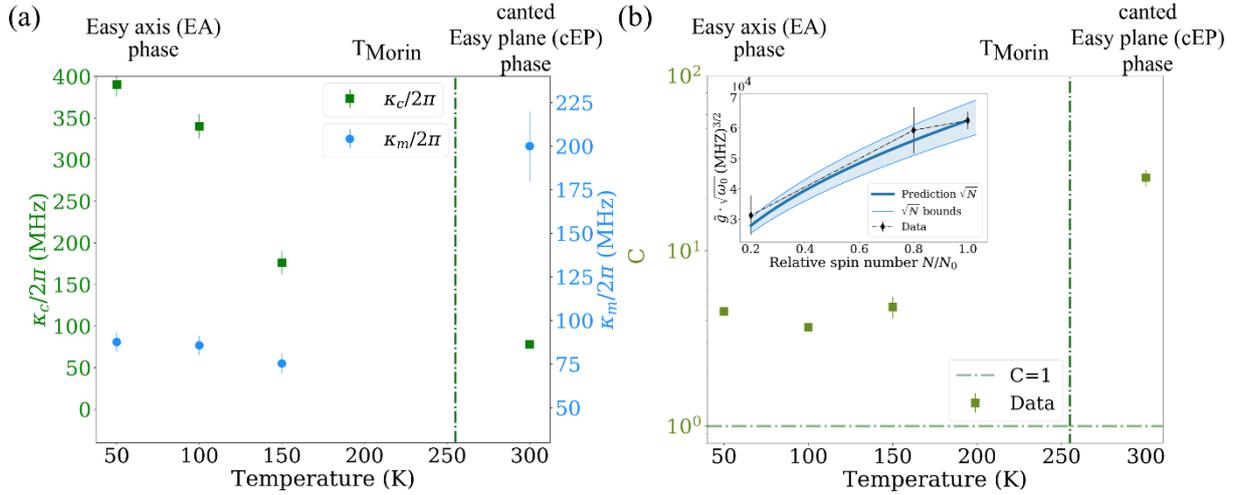

*Figure 3: **Coherence of the hybrid states.** (a) Temperature dependence of the cavity resonator's (green) and magnon linewidth (blue). The magnon linewidth decreases towards $T_{Morin}$, in line with previous observations ([43]) (b) Temperature dependence of the cooperativity C in the colinear easy-axis phase and the canted easy-plane phase of hematite. C > 1 is observed at all temperatures, indicating a coherent exchange of information. Inset: Coupling strength dependency for different sample sizes at 300 K. Experimental coupling strengths (black, dotted line) are in agreement with the theoretical prediction using an upper and a lower bound. The VNA output power level was set to -17 dBm.*

In summary, we demonstrate the strong coupling regime between microwave cavity field and collective spin excitations in both collinear and non-collinear antiferromagnets over a wide temperature range. Theoretically and experimentally, we highlight that the coupling strength scales with the materials' anisotropy, exchange and DMI fields. Altogether, despite the small net moments, our results highlight the potential of using insulating antiferromagnets with different magnetic symmetries for cavity spintronics. Taking advantages of the ultrafast response of antiferromagnets, we can envision possible applications ranging from high frequency sensing, non-reciprocal ICT devices [18], to building blocks for quantum information processing if other nonlinear elements such as qubits are added. Finally, the prospects of much higher accessible resonance frequencies for antiferromagnetic cavity spintronics open new realms for integrating A-CMPs to novel spintronic architectures.

**Acknowledgements:** R.L., I. B., A.A. and M.K. acknowledge financial support from the Horizon 2020 Framework Programme of the European Commission under FET-Open grant agreement no. 863155 (s-Nebula). M.K. acknowledges support from the Graduate School of Excellence Materials Science in Mainz (MAINZ) DFG 266, the DAAD (Spintronics network, Project No. 57334897). M.K. acknowledges support from the DFG project number 423441604 and 268565370 (SFB TRR 173 Spin+X Project A01). The Research Council of Norway supported H.T.S., B.B, M.K., and A.B. through its Centres of Excellence funding scheme, project number 262633 "QuSpin".

## Bibliography

[1] H. Huebl *et al.*, 'High Cooperativity in Coupled Microwave Resonator Ferrimagnetic Insulator Hybrids', *Physical Review Letters*, vol. 111, no. 12, p. 127003, 2013, doi: 10.1103/physrevlett.111.127003.

[2] Y. Tabuchi, S. Ishino, T. Ishikawa, R. Yamazaki, K. Usami, and Y. Nakamura, 'Hybridizing Ferromagnetic Magnons and Microwave Photons in the Quantum Limit', *Physical Review Letters*, vol. 113, no. 8, p. 083603, 2014, doi: 10.1103/physrevlett.113.083603.

[3] X. Zhang, C.-L. Zou, L. Jiang, and H. X. Tang, 'Strongly Coupled Magnons and Cavity Microwave Photons', *Phys. Rev. Lett.*, vol. 113, no. 15, Oct. 2014, doi: 10.1103/physrevlett.113.156401.

[4] X. Zhang, C.-L. Zou, N. Zhu, F. Marquardt, L. Jiang, and H. X. Tang, 'Magnon dark modes and gradient memory', *Nat. Commun.*, vol. 6, no. 1, Nov. 2015, doi: 10.1038/ncomms9914.

[5] M. Harder *et al.*, 'Level Attraction Due to Dissipative Magnon-Photon Coupling', *Phys. Rev. Lett.*, vol. 121, no. 13, Sep. 2018, doi: 10.1103/physrevlett.121.137203.

[6] M. Harder, B. M. Yao, Y. S. Gui, and C.-M. Hu, 'Coherent and dissipative cavity magnonics', *Journal of Applied Physics*, vol. 129, no. 20, p. 201101, May 2021, doi: 10.1063/5.0046202.

[7] A. V. Chumak and H. Schultheiss, 'Magnonics: spin waves connecting charges, spins and photons', *Journal of Physics D: Applied Physics*, vol. 50, no. 30, p. 300201, 2017, doi: 10.1088/1361-6463/aa7715.

[8] Y. Tabuchi, S. Ishino, A. Noguchi, T. Ishikawa, R. Yamazaki, K. Usami, Y. Nakamura, 'Coherent coupling between a ferromagnetic magnon and a superconducting qubit', *Science*, vol. 349, no. 6246, pp. 405–408, Jul. 2015, doi: 10.1126/science.aaa3693.

[9] D. Lachance-Quirion *et al.*, 'Resolving quanta of collective spin excitations in a millimeter-sized ferromagnet', *Science Advances*, vol. 3, no. 7, p. e1603150, 2017, doi: 10.1126/sciadv.1603150.

[10] D. Lachance-Quirion, S. P. Wolski, Y. Tabuchi, S. Kono, K. Usami, and Y. Nakamura, 'Entanglement-based single-shot detection of a single magnon with a superconducting qubit', *Science*, vol. 367, no. 6476, pp. 425–428, Jan. 2020, doi: 10.1126/science.aaz9236.

[11] S. P. Wolski *et al.*, 'Dissipation-Based Quantum Sensing of Magnons with a Superconducting Qubit', *Phys. Rev. Lett.*, vol. 125, no. 11, p. 117701, Sep. 2020, doi: 10.1103/PhysRevLett.125.117701.

[12] L. Bai *et al.*, 'Cavity Mediated Manipulation of Distant Spin Currents Using a Cavity-Magnon-Polariton', *Physical Review Letters*, vol. 118, no. 21, p. 217201, 2017, doi: 10.1103/physrevlett.118.217201.

[13] M. Aspelmeyer, T. J. Kippenberg, and F. Marquardt, 'Cavity optomechanics', *Rev. Mod. Phys.*, vol. 86, no. 4, pp. 1391–1452, Dec. 2014, doi: 10.1103/revmodphys.86.1391.

[14] S. V. Kusminskiy, H. X. Tang, and F. Marquardt, 'Coupled spin-light dynamics in cavity optomagnonics', *Physical Review A*, vol. 94, no. 3, p. 033821, Sep. 2016, doi: 10.1103/physreva.94.033821.

[15] T. S. Parvini, V. A. S. V. Bittencourt, and S. V. Kusminskiy, 'Antiferromagnetic cavity optomagnonics', *Phys. Rev. Research*, vol. 2, no. 2, p. 022027, May 2020, doi: 10.1103/PhysRevResearch.2.022027.

[16] D. Lachance-Quirion, Y. Tabuchi, A. Gloppe, K. Usami, and Y. Nakamura, 'Hybrid quantum systems based on magnonics', *Applied Physics Express*, vol. 12, no. 7, p. 070101, Jun. 2019, doi: 10.7567/1882-0786/ab248d.

[17] D. Zhang, X.-Q. Luo, Y.-P. Wang, T.-F. Li, and J. Q. You, 'Observation of the exceptional point in cavity magnon-polaritons', *Nat. Commun.*, vol. 8, no. 1, Nov. 2017, doi: 10.1038/s41467-017-01634-w.

[18] J. W. Rao, Y. T. Zhao, Y. S. Gui, X. L. Fan, D. S. Xue, and C.-M. Hu, 'Controlling Microwaves in Non-Hermitian Metamaterials', *Phys. Rev. Applied*, vol. 15, no. 2, p. L021003, Feb. 2021, doi: 10.1103/PhysRevApplied.15.L021003.

[19] I. Boventer and C. Dörflinger and T. Wolz and R. Macedo and R. Lebrun and M. Kläui and M. Weides, 'Control of the coupling strength and linewidth of a cavity magnon-polariton', *Physical Review Research*, vol. 2, no. 1, Feb. 2020, doi: 10.1103/physrevresearch.2.013154.

[20] Y.-P. Wang and C.-M. Hu, 'Dissipative couplings in cavity magnonics', *Journal of Applied Physics*, vol. 127, no. 13, p. 130901, Apr. 2020, doi: 10.1063/1.5144202.

[21] P.-C. Xu, J. W. Rao, Y. S. Gui, X. Jin, and C.-M. Hu, 'Cavity-mediated dissipative coupling of distant magnetic moments: Theory and experiment', *Physical Review B*, vol. 100, no. 9, Sep. 2019, doi: 10.1103/physrevb.100.094415.

[22] Y. Yang and J. W. Rao and Y. S. Gui and B. M. Yao and W. Lu and C. -M. Hu, 'Control of the magnon-

photon level attraction in a planar cavity', 2019.

[23] V. Cherepanov, I. Kolokolov and V. Lvov, 'The saga of YIG: Spectra, thermodynamics, interaction and relaxation of magnons in a complex magnet', *Phys. Rep.*, vol. 229, no. 3, pp. 81–144, 1993, doi: http://dx.doi.org/10.1016/0370-1573(93)90107-O.

[24] C. Hauser *et al.*, 'Yttrium Iron Garnet Thin Films with Very Low Damping Obtained by Recrystallization of Amorphous Material', *Scientific Reports*, vol. 6, no. 1, Feb. 2016, doi: 10.1038/srep20827.

[25] C. Dubs *et al.*, 'Low damping and microstructural perfection of sub-40nm-thin yttrium iron garnet films grown by liquid phase epitaxy', *Physical Review Materials*, vol. 4, no. 2, Feb. 2020, doi: 10.1103/physrevmaterials.4.024416.

[26] T. Jungwirth, J. Sinova, A. Manchon, X. Marti, J. Wunderlich, and C. Felser, 'The multiple directions of antiferromagnetic spintronics', *Nature Physics*, vol. 14, no. 3, pp. 200–203, Mar. 2018, doi: 10.1038/s41567-018-0063-6.

[27] S. M. Rezende, A. Azevedo, and R. L. Rodríguez-Suárez, 'Introduction to antiferromagnetic magnons', *Journal of Applied Physics*, vol. 126, no. 15, p. 151101, Oct. 2019, doi: 10.1063/1.5109132.

[28] V. Baltz, A. Manchon, M. Tsoi, T. Moriyama, T. Ono, and Y. Tserkovnyak, 'Antiferromagnetic spintronics', *Reviews of Modern Physics*, vol. 90, no. 1, p. 015005, 2018, doi: 10.1103/revmodphys.90.015005.

[29] R. Lebrun *et al.*, 'Tunable long-distance spin transport in a crystalline antiferromagnetic iron oxide', *Nature*, vol. 561, no. 7722, pp. 222–225, Sep. 2018, doi: 10.1038/s41586-018-0490-7.

[30] R. Lebrun *et al.*, 'Long-distance spin-transport across the Morin phase transition up to room temperature in ultra-low damping single crystals of the antiferromagnet α-$Fe_2O_3$', *Nat Commun*, vol. 11, no. 1, p. 6332, Dec. 2020, doi: 10.1038/s41467-020-20155-7.

[31] I. Boventer, H. T. Simensen, A. Anane, M. Kläui, A. Brataas, and R. Lebrun, 'Room-Temperature Antiferromagnetic Resonance and Inverse Spin-Hall Voltage in Canted Antiferromagnets', *Phys. Rev. Lett.*, vol. 126, no. 18, p. 187201, May 2021, doi: 10.1103/PhysRevLett.126.187201.

[32] J. Li *et al.*, 'Spin current from sub-terahertz-generated antiferromagnetic magnons', *Nature*, vol. 578, no. 7793, pp. 70–74, Feb. 2020, doi: 10.1038/s41586-020-1950-4.

[33] P. Vaidya *et al.*, 'Subterahertz spin pumping from an insulating antiferromagnet', *Science*, vol. 368, no. 6487, pp. 160–165, Apr. 2020, doi: 10.1126/science.aaz4247.

[34] M. Mergenthaler *et al.*, 'Strong coupling of microwave photons to antiferromagnetic fluctuations in an organic magnet', *Physical Review Letters*, vol. 119, no. 14, p. 147701, 2017, doi: 10.1103/PhysRevLett.119.147701.

[35] M. Białek, J. Zhang, H. Yu, and J.-Ph. Ansermet, 'Strong Coupling of Antiferromagnetic Resonance with Subterahertz Cavity Fields', *Phys. Rev. Applied*, vol. 15, no. 4, p. 044018, Apr. 2021, doi: 10.1103/PhysRevApplied.15.044018.

[36] K. Grishunin *et al.*, 'Terahertz Magnon-Polaritons in $TmFeO_3$', *ACS Photonics*, vol. 5, no. 4, pp. 1375–1380, Apr. 2018, doi: 10.1021/acsphotonics.7b01402.

[37] P. Sivarajah *et al.*, 'THz-frequency magnon-phonon-polaritons in the collective strong-coupling regime', *Journal of Applied Physics*, vol. 125, no. 21, p. 213103, Jun. 2019, doi: 10.1063/1.5083849.

[38] M. Bialek, A. Magrez, and J.-P. Ansermet, 'Spin-wave coupling to electromagnetic cavity fields in dysposium ferrite', *Physical Review B*, vol. 101, no. 2, Jan. 2020, doi: 10.1103/physrevb.101.024405.

[39] Ø. Johansen and A. Brataas, 'Nonlocal Coupling between Antiferromagnets and Ferromagnets in Cavities', *Physical Review Letters*, vol. 121, no. 8, Aug. 2018, doi: 10.1103/physrevlett.121.087204.

[40] I. Boventer *et al.*, 'Supplementary Material: Strongly coupled antiferromagnetic magnons and cavity photons in collinear and canted phases of hematite'.

[41] H. J. Fink, 'Resonance Line Shapes of Weak Ferromagnets of the α-$Fe_2O_3$ and $NiF_2$ Type', *Physical Review*, vol. 133, no. 5A, pp. A1322–A1326, Mar. 1964, doi: 10.1103/physrev.133.a1322.

[42] F. J. Morin, 'Electrical Properties of α-Fe2O3 and α-Fe2O3 Containing Titanium', *Physical Review*, vol. 83, no. 5, pp. 1005–1010, 1951, doi: 10.1103/physrev.83.1005.

[43] R. Lebrun *et al.*, 'Anisotropies and magnetic phase transitions in insulating antiferromagnets determined by a Spin-Hall magnetoresistance probe', *Commun Phys*, vol. 2, no. 1, p. 50, Dec. 2019, doi: 10.1038/s42005-019-0150-8.

[44] C. E. Johnson, L. A. Prelorendjo, and M. F. Thomas, 'Field induced spin reorientation in orthoferrites DyFeO3, HoFeO3 and ErFeO3', *Journal of Magnetism and Magnetic Materials*, vol. 15–18, pp. 557–558, Jan. 1980, doi: 10.1016/0304-8853(80)90662-9.

[45] A. V. Kimel *et al.*, 'Optical excitation of antiferromagnetic resonance in $TmFeO_3$', *Phys. Rev. B*, vol. 74, no. 6, p. 060403, Aug. 2006, doi: 10.1103/PhysRevB.74.060403.

[46] H. Watanabe, T. Kurihara, T. Kato, K. Yamaguchi, and T. Suemoto, 'Observation of long-lived coherent spin precession in orthoferrite $ErFeO_3$ induced by terahertz magnetic fields', *Appl. Phys. Lett.*, vol. 111, no. 9, p. 092401, Aug. 2017, doi: 10.1063/1.4985035.

[47] V. M. Judin, A. B. Sherman, and I. E. Myl'nikova, 'Magnetic properties of $YFeO_3$', *Physics Letters*, vol. 22, no. 5, pp. 554–555, Sep. 1966, doi: 10.1016/0031-9163(66)90649-4.

[48] Z. Jin, Z. Mics, G. Ma, Z. Cheng, M. Bonn, and D. Turchinovich, 'Single-pulse terahertz coherent control of spin resonance in the canted antiferromagnet YFeO$_3$, mediated by dielectric anisotropy', *Phys. Rev. B*, vol. 87, no. 9, p. 094422, Mar. 2013, doi: 10.1103/PhysRevB.87.094422.

[49] K. P. Belov, A. M. Kadomtseva, N. M. Kovtun, V. N. Derkachenko, V. N. Melov, and V. A. Khokhlov, 'On the character of phase transitions in ErFeO$_3$', *Phys. Stat. Sol. (a)*, vol. 36, no. 1, pp. 415–425, Jul. 1976, doi: 10.1002/pssa.2210360145.